\documentclass[preprint2]{aastex}

\shorttitle{Vrba et al.}
\shortauthors{Embedded Cluster Near SGR 1900+14}

\begin{document}


\title{The Discovery of an \\
  Embedded Cluster of High-Mass Stars Near SGR 1900+14 }

\author{Frederick J. Vrba, Arne A. Henden\altaffilmark{1}, Christian B. Luginbuhl, and Harry H. Guetter}
\affil{U.S. Naval Observatory, Flagstaff Station,
    Flagstaff, AZ 86002-1149}

\author{Dieter H. Hartmann}
\affil{Dept. of Physics and Astronomy,
    Clemson, University, Clemson, SC 29634-0978}

\and

\author{Sylvio Klose}
\affil{Th\"uringer Landessternwarte Tautenburg, D--07778 Tautenburg, Germany}


\altaffiltext{1}{Universities Space Research Association}


\begin{abstract}

Deep I-band imaging to I~$\approx$~26.5 of the soft gamma--ray
repeater SGR 1900+14 region has revealed a compact cluster of massive
stars located only a few arcseconds from the fading radio
source thought to be the location of the SGR \citep{fra99}. This
cluster was previously hidden in the glare of the pair of M5 supergiant 
stars (whose light was removed by PSF subtraction) proposed by 
\citet{vrb96} as likely associated with the SGR 1900+14. The cluster
has at least 13 members within a cluster radius of $\approx$~0.6~pc
based on an estimated distance of 12--15 kpc. It is remarkably
similar to a cluster found associated with SGR 1806--20 \citep{fuc99}.
That similar clusters have now been found at or near the
positions of the two best--studied SGRs suggests that young neutron
stars, thought to be responsible for the SGR phenomenon, have their
origins in proximate compact clusters of massive stars.

\end{abstract}


\keywords{gamma rays: bursts -- gamma rays: observations-- star clusters}


\section{Introduction}

The \citet[V96]{har96,vrb96} survey of the original Network 
Synthesis Localization (NSL) of SGR 1900+14 \citep{hur94} found a pair 
of nearly identical M5 supergiant stars, separated by 3.3 arcsec, and 
at an estimated distance of 12-15 kpc. While just outside of the original 
NSL, they lie within the ROSAT HRI localization of the quiescent
X--ray source RX~J190717+0919.3 thought to be associated with SGR 1900+14 
\citep{hur96}. On the basis of the small probability that even one supergiant 
would lie within the ROSAT error circle and that at least one other 
supergiant had been associated with an SGR (1806--20; \citet{van95,kul95}), 
V96 proposed that the M star pair may be associated with the SGR 1900+14 
source. The position of the M star pair has continued to be consistent with more
recent X--ray and gamma--ray observations which, taken together, have narrowed
considerably the actual location of SGR 1900+14 from the original NSL 
area of 5 arcmin$^2$.  These recent X--ray and gamma--ray observations
have also detected variations with a period of 5.16 sec
\citep{hur99a,mur99,kou99} and a deceleration of $\dot P \approx
10^{-10}$~sec/sec.  Taken together, these are interpreted as
evidence that the SGR source is a magnetar, though there remains
some uncertainty in this interpretation \citep{mar99}.

Additionally, a variable and fading radio source was detected 
shortly after the 27 August SGR 1900+14 superburst by \citet{fra99}, providing 
strong evidence that it was the radio counterpart to the SGR. Its
subarcsec accurate position is located only a few arcseconds from the M
stars. These positional coincidences, the lack of a plerionic radio source, 
and, despite arguments for SNR G42.8+0.6 in the literature, the lack of a 
coincident supernova remnant, suggest that the system of proximate, 
high--mass M stars should not yet be dismissed as an evolutionary companion 
to the pulsating X--ray source associated with the SGR. 

Finding direct evidence that the M star pair may be associated with SGR
1900+14 has proven elusive as summarized by \citet{gue20}. Also difficult
is a theoretical understanding of how isolated, albeit high mass, stars
could play a role in the formation of a pulsating X--ray source, despite the
presence of a high mass luminous blue variable (LBV) very near the SGR 1806--20
localization position, a remarkably similar situation to that for SGR
1900+14. Recent near-- and mid--infrared observations of SGR 1806--20
\citep[F99]{fuc99}, however, have revealed the LBV to be only the most luminous 
member of a compact cluster of massive stars. Such proximate regions 
of recent star formation provide a natural location for the birth of 
such pulsating X--ray sources,
which cannot be very old, without the need for invoking
enormous space velocities from the nearest supernova remnants.

In this paper we present evidence for a similar compact cluster of
high-mass stars which has heretofore been hidden in the
glare of its brightest components, the pair of M5 supergiant stars.

\section{Observations}

The 1998 outburst season of SGR 1900+14 presented an opportunity
to search for optical and near-infrared
variability of the double M stars, or other sources
within the ROSAT HRI error circle, which might be correlated to the 
SGR outbursts via some process such as mass transfer to a compact object. 
Beginning in early May and continuing through mid--July 1998 we
carried out an I-- and J--band monitoring campaign at the U.S. Naval
Observatory,
Flagstaff Station (NOFS) which eventually comprised 2025 short exposure frames 
of data with 54,460 seconds of open shutter time during 16 nights,
intended to sample variability timescales down to a few seconds. The
results of this work found no variablity for any object within
the ROSAT HRI error circle and are presented more fully in \citet{vrb20}.
 
However, it was recognized that the numerous short I--band exposures
constituted several hours of total exposure time, which could be
stacked to form a deep I--band image to search for a counterpart at
the position of the \citet{fra99} variable radio
source. To the 1998 data were added additional short exposure frames
from 1995 and 1999. In all, 217 frames of individual exposure time
between 1 and 10 minutes were coadded to form a net image of about
6.5 hours total exposure. All frames were obtained with one
of two Tektronix 2K CCDs on the 1.55--m Strand Astrometric Telescope at the
USNOFS. It was additionally
recognized that, since the exposures used were short enough not to
saturate the three bright M stars (A, B,and C of V96), their
light could largely be removed by PSF subtraction.  

Figure~1 is an approximately 45 x 45 arcsec portion of the median--filtered
composite I--band image centered on the V96 M stars, with a limiting 
detection magnitude of I~$\approx$~26.5. In this image the M
stars ABC have been removed, although their positions are still
apparent due to imperfect subtraction. 
The position of the variable radio source is shown, but
no counterpart is visible to I~$\approx$~26.5, which is consistent with the
non-detections in the near infrared of \citet{eik99}. Unfortunately,
none of the nearly 200 frames from 1998 were obtained simultaneously
with a gamma--ray burst from SGR 1900+14.

Of greater interest is that the subtracted I--band image shows what appears 
to be a cluster of stars, and possibly nebulosity, centered on the position 
of the M stars. The IRAS source found at this location by
\citet{van96} shows a steeply rising energy spectrum that can be
interpreted as warm dust in the cluster region.
Figure~1 also shows
identification numbers of the possible cluster stars. On UT 1999 October 28 
we used the ASTROCAM IR imager, which employs an SBRC 1024$^2$ InSb 
detector, at the 1.55--m telescope to obtain a 1600 second net exposure 
J--band image of this region. An approximately 45 x 45 arcsec region
of this image is shown in Figure~2, where again the M stars were
somewhat successfully PSF--subtracted.

We obtained photometry for the cluster stars from the I-- and J--band frames,
calibrated with several I-- and J--band local standards which had previously
been set up for our variability monitoring program. The photometric
results are presented in Table~1 where the results for stars 5 and 6
are presented together as they could not be separated in the J--band
observations. The observed (I--J)~$\approx$~7 colors are far larger
than for any unreddened star and indicate that they suffer extremely high 
extinction.

\section{Nature of the Cluster}

Assuming that the cluster stars are at the
same distance and suffer the same extinction as the M supergiant stars 
(12-15 kpc; A$_V$~=~19.2$\pm$1.0; V96) we placed all stars in an
$M_I$ vs. (I--J) CM diagram (Figure~3), assuming normal interstellar
extinction \citep{bes88}, and  where the error bars include the
ranges in distance and A$_V$ values given above. The solid curves 
show the approximate loci for supergiants and dwarfs later than A0 and for
giants later than G0 while the dashed lines show the M0 (I--J) colors,
for reference. The large uncertainty of the intrinsic (I--J) colors of the
stars after subtracting a huge baseline of extinction renders them
essentially useless in estimating their spectral types. However, at
this assumed distance and extinction
the stars have luminosities far greater than
that of main sequence stars. We note that even assuming the stars
are at a much closer distance (for instance d~$\approx$~5 kpc as has
often been quoted by association with the SNR G42.8+0.6) has little affect
on the conclusion that these are highly luminous stars.

Several examples of compact high mass young clusters serve as templates for
these objects: NGC 3603 \citep{mof94}, W43 \citep{blu99}, and several
clusters summarized in \citet{fig99}. These clusters are characterized
by 10 -- 30 cluster members, radii of 0.2 -- 1.0 pc, and ages 
of 1 -- 10 Myr. The SGR 1900+14 cluster has at least 13 members
(including stars A
and B) and an approximate 7 arcsec radius which,
at a distance of 12 -- 15 kpc, corresponds to a cluster radius of
$\approx$ 0.4 pc. A remarkably similar example to that of the SGR 1900+14
cluster is described by \citet{mof76} in which a group of 12 luminous stars
surround the M3 I supergiant star HD143183 within a cluster radius of
0.6~pc. These examples support the idea that the small cluster of stars
near SGR 1900+14 and dominated by the M5 supergiants is likely a real
association. A formal astrometric solution, not previously presented, for 
the positions of the M supergiants based on 21 USNO-A2.0 stars gives the 
result ($\pm$ 0.1 arcsec):

\centerline{Star A: $\alpha$~=~19$^h$~07$^m$~15.35$^s$, $\delta$~=~+09$^d$~19'~21.4" (J2000)}

\centerline{Star B: $\alpha$~=~19$^h$~07$^m$~15.13$^s$, $\delta$~=~+09$^d$~19'~20.7" (J2000)}

\section{Discussion}

If the cluster was the birthplace of SGR 1900+14, this essentially excludes
SNR G42.8+0.6 as playing any role in the SGR. Although one can envision
scenarios in which the SNR progenitor was ejected from the cluster by
dynamical interaction or a much earlier supernova, this leaves the
necessity of the neutron star having been kicked back to almost exactly 
its place of origin by the supernova that formed SNR G42.8+0.6 (since the 
cluster and SGR localizations are coincident), an unlikely coincidence both 
in space and timing. However, despite the association of G42.8+0.6 with
SGR 1900+14 in the literature, there has been no  evidence supporting this 
association offered, such as the probablity of finding any SNR within a
given distance, based on the number density of SNRs in the Galactic plane.

A more plausable scenario is one in which the cluster and associated
dense gas/dust cloud hides a recent supernova. Evidence for this
cloud comes from Figures 1 and 2 and the coincident extended strong 
far--infrared source indicating compact warm and extended cool dust
(see V96). Optical extinction from this cloud combined with a 12--15~kpc
distance explains why the supernova would not have been noticed
historically. A very young SNR expanding into the dense wind--blown
bubble due to mass loss from the supergiant stars in the cluster would
be consistent with the otherwise unexplained persistent X--ray source at 
this position, RX J190717+0919.3 \citep{hur96}. While no quiescent radio source
is known at this position, a combination of self--absorption within
the dense medium and rapid decay \citep{rey88} could account for this.
The supernova remnant evolutionary calculations of \citet{tru99} indicate that 
for an ejecta mass of 1 M$_\sun$, and an external density medium of
10 cm$^{-3}$, one finds characteristic sizes of $\approx$ 1~pc at
t~=~1000 yr; similar to that of the cluster dimensions at the estimated
M supergiant distances.

The most likely position for the SGR itself is the \citet{fra99} 
fading radio source located at
$\alpha$~=~19$^h$~07$^m$~14.33$^s$, $\delta$~=~+09$^d$~19'~21.1"
(J2000), with positional accuracy of $\pm$ 0.15 arcsec in each coordinate. 
With these astrometric positions we estimate the approximate distances from 
the center and edge of the cluster to the radio position as 12 arcsec
(0.7--0.9~pc) and 5 arcsec (0.3--0.4~pc), respectively, based on the 12--15~kpc 
distance estimate.
Thus, even at the extreme minimum age of the SGR based on the simplest 
magnetar physics ($\approx$~700 yr; Kouveliotou et al. 1999) this implies 
a tangential velocity of $\approx$ 420 km~s$^{-1}$ from the near edge of the
cluster. While still an ample velocity for the runaway neutron star,
it obviates the enormous space velocities implied by associating it with 
G42.8+0.6 \citep{kou99}, which is about 12 arcmin away \citep{hur99b}.

While an isolated instance of the compact, high mass cluster found
at/near SGR 1900+14 would be dismissed as a chance superposition,
its striking similarity to the cluster found near SGR 1806-20 by \citet{fuc99}
must be recognized. In that case,
an LBV supergiant is found associated with a cluster of at least another 
four massive young stars enshrouded in a bright dust cloud as imaged by
ISO and located only 7 arcsec from the SGR gamma--ray localization. 
With an approximate cluster radius of 8 arcsec and an estimated
distance of 14.5 kpc, this implies a cluster radius of $\approx$~0.6~pc.
Now that similar compact clusters have been found near the 
positions of the two best studied SGRs (1806-20 and 1900+14)
the possiblity that young SGR neutron stars have their origins in
compact clusters should be considered seriously.

\clearpage



\figcaption[fig1.ps]{An approximately 45 x 45 arcsec portion of the
6.5 hour exposure I--band image of the SGR 1900+14 region formed from
numerous short exposures as explained in the text. North is at the
top, East to the left. In this image the bright M stars, discussed in
\citet{vrb96}, have been subtracted revealing the cluster of faint
stars. The position of the \citet{fra99} fading radio source is
indicated by the circle. The 11 stars forming the cluster are
numbered for identification. \label{fig1}}

\figcaption[fig2.ps]{An approximately 45 x 45 arcsec portion of a
J-band image of the SGR 1900+14 region. North is at the top and East
to the left. In this image the bright M stars, discussed in \citet{vrb96}, 
have been subtracted, showing the nebulosity associated with the
cluster. \label{fig2}}

\figcaption[fig3.ps]{The M$_I$ vs I--J color--magnitude diagram for
the 11 newly discovered cluster stars. The I and J band photometry
has been dereddened by A$_V$ = 19.2 $\pm$ 1.0 magnitudes and the
cluster stars have been assumed to be at a distance range of 12 to
15~kpc as explained in the text. The approximate loci for luminosity 
class I, III, and V stars are also shown by the solid curves, with the
positions for M0 spectral type stars shown for reference by the dashed
lines. \label{fig3}}





\clearpage

\begin{deluxetable}{cccc}
\footnotesize
\tablecaption{I-- and J--Band Photometry of Cluster Stars \label{tbl-1}}
\tablewidth{0pt}
\tablehead{
\colhead{Star} & \colhead{I $\pm$ $\sigma$(I)}   & \colhead{J $\pm$ $\sigma$(J)}   & \colhead{(I -- J) $\pm$ $\sigma$(I -- J)} 
}
\startdata
1 &  20.00 $\pm$ 0.02 & 13.17 $\pm$  0.02 & 6.83 $\pm$ 0.03 \\
2 &  21.10 $\pm$ 0.02 & 14.26 $\pm$  0.02 & 6.84 $\pm$ 0.03 \\
3 &  21.85 $\pm$ 0.02 & 14.51 $\pm$  0.02 & 7.34 $\pm$ 0.03 \\
4 &  22.01 $\pm$ 0.02 & 15.06 $\pm$  0.04 & 6.95 $\pm$ 0.05 \\
5 &  22.77 $\pm$ 0.03 & \nodata           &  \nodata        \\
6 &  23.18 $\pm$ 0.04 & \nodata           &  \nodata        \\
5+6& 22.20 $\pm$ 0.04 & 15.47 $\pm$  0.06 & 6.73 $\pm$ 0.07 \\
7 &  23.16 $\pm$ 0.05 & 15.70 $\pm$  0.09 & 7.46 $\pm$ 0.10 \\
8 &  22.85 $\pm$ 0.04 & 15.62 $\pm$  0.06 & 7.23 $\pm$ 0.07 \\
9 &  23.01 $\pm$ 0.05 & 15.53 $\pm$  0.05 & 7.48 $\pm$ 0.07 \\
10&  23.60 $\pm$ 0.06 & 16.41 $\pm$  0.07 & 7.19 $\pm$ 0.09 \\
11&  23.78 $\pm$ 0.06 & 16.12 $\pm$  0.04 & 7.66 $\pm$ 0.07 \\
 \enddata
\end{deluxetable}

\end{document}